# Causal inference in drug discovery and development

Tom Michoel[1] and Jitao David Zhang[2,3]

[1] Computational Biology Unit, Department of Informatics, University of Bergen, Postboks 7803, 5020 Bergen, Norway
[2] Pharma Early Research and Development, Roche Innovation Centre Basel, F. Hoffmann-La Roche, Grenzacherstrasse 124, 4070 Basel, Switzerland
[3] Department of Mathematics and Computer Science, University of Basel, Spiegelgasse 1, 4051 Basel, Switzerland

## Teaser

To explain or to predict? To understand.

## Abstract

To discover new drugs is to seek and to prove causality. As an emerging approach leveraging human knowledge and creativity, data, and machine intelligence, causal inference holds the promise of reducing cognitive bias and improving decision making in drug discovery. While it has been applied across the value chain, the concepts and practice of causal inference remain obscure to many practitioners. This article offers a non-technical introduction to causal inference, reviews its recent applications, and discusses opportunities and challenges of adopting the causal language in drug discovery and development.

## Keywords

## Main Text

## 1. Introduction

Causal inference is the process of identifying causal effects based on prior knowledge, hypothesis, and correlations observed in data. This article aims at equipping practitioners of drug discovery and development, especially those without formal training in mathematics and statistics, with necessary knowledge to start working with causal inference at three levels: (1) to recognize situations where causal inference is advantageous and to understand why a causal model is needed (Section 2), (2) to perform causal inference step by step with software (Section 3), and (3) to learn and get inspiration from recent applications (Section 4). Finally, in Section 5, we discuss opportunities and challenges of adopting the causal language in drug discovery and development.

Causal inference identifies causations from correlations. In statistics, correlation means the relationship between two random variables, whether causal or not. For instance, high temperatures lead to higher ice-cream sales, therefore the variables *temperature* and *ice-cream sales* are correlated. Similarly, high temperatures make wildfire more likely to happen, therefore the variables *temperature* and *wildfire frequency* are correlated. Apparently, *ice-cream sales* and *wildfire frequency* are correlated but do not cause each other. The task of causal inference in this context is to infer the causal relationship between *temperature, ice-cream sales,* and *wildfire frequency* based on observations.

There is neither a consensus definition of causality nor a single way to identify it (Box 1). It is partially because the nature of causation is still under debate[1], and partially because even simple causality like *higher temperature causes higher ice-cream sales* can be broken down into an infinite chain of causal relationships that can involve entities across both physical (molecules, receptors, cells, organs and systems, society, *etc.*) and time scales.

In this article, we introduce the statistical causal inference approach, and define causality as a probabilistic relationship that satisfies four conditions: *regular probabilistic update*, *manipulation*, *counterfactual condition*, and *mechanism of action*.

Take the example of testing whether a drug is causal for halting deadly cancer progression in a clinical trial. *Regular probabilistic update* means that taking the drug modifies the conditional probability of dying from the disease within a defined time window, irrespective of where and when the trial happens. *Manipulation* means that drug treatment shows additional benefit even if we consider all other factors affecting patients' survival, for instance age and comorbidities. *Counterfactual condition* means that the death of a patient would not have been postponed had the drug not been taken. Finally, *mechanism of action* means that we understand why the drug prolongs patients' survival, for instance by activating tumour-infiltrating immune cells. Taken together, the four conditions ensure both statistical correlation and mechanistic understanding. They put the causality criteria by Austin Bradford Hill[2] in a causal context, and implement a practical test for the philosophical reasoning of establishing causality in healthcare.[3]

# 2. What is causal inference and why do we need it

## 2.1 Why causality matters in drug discovery and development

Causality is indispensable for predicting outcomes of intervention, for answering counterfactual, *what-if* questions, and for human understanding. Although it takes time and effort to identify causes from correlations, the investment is rewarding. The history of drug discovery is bound with cases where causality inspires repurposing of a drug, developing new classes of chemical matters, or refuting the use of a drug.

One prominent example is the Shakespearean history of thalidomide[4]. Chemically, thalidomide is a mixture of two enantiomers: the (R)-enantiomer is sedative while the (S)-enantiomer is teratogenic. While its use was first correlated with an unusually high incidence of birth deficiency[5], it took much time and work to identify the cause for the teratogenicity, *i.e.* the degradation of SALL4 protein via E3 ubiquitin ligase complex[6]. In the meantime, thalidomide was found potentially effective against drug-resistant multiple myeloma. Follow-up study based on the first causal finding offered a causal explanation for this correlation, too: thalidomide induces protein degradation of key transcription factors *Ikaros* (IKZF1) and *Aiolos* (IKZF3) with the same ubiquitin apparatus. Last but not least, the causal link between a small molecule and protein degradation inspired a flourishing search for new modalities including proteolysis-targeting chimera (PROTAC)[7] and other multispecific drugs[8].

Repurposing drugs to treat COVID-19 offers another recent example demonstrating the importance of causality. Much literature reported a correlation between treatment with drugs targeting one or both human sigma receptors ($\sigma_1$ and $\sigma_2$), among others hydroxychloroquine, and negative modulation of SARS-Cov-2 infection. Researchers, however, noticed the intriguing discrepancy that while most compounds have nanomolar affinity against sigma receptors, the cellular antiviral activities showed large variation. It raised the question whether pharmacological modulation of sigma receptors is causal for SARS-Cov-2 inhibition. The question led *Tummino* et al. to find that phospholipidosis, instead of target-mediated mechanisms, underlies the antiviral activity of many drugs[9]. Phospholipidosis is caused by physicochemical properties of the drugs, can be reliably predicted experimentally and computationally[10], and results in dose-limiting toxicity. The example, opposite to the example of thalidomide, highlights the importance of dissecting causal mechanisms from correlations, too.

While correlations suffice for predictions, causality enables us to understand why a drug works or causes harm. What is the relationship between correlation and causation? And how can we systematically identify causations from correlations?

## 2.2 Distinguishing causation from correlation

Causality manifests itself by correlations, though correlations are not always caused by causality. There are four reasons why we observe a correlation between two events: *causation*, *confounding*, *coincidence*, and *conspiracy*.

We assume that two variables, *x* and *y*, depict expression levels of two proteins, X and Y, in a population of cells. If *x* and *y* is correlated, four scenarios are possible:

1. *Causation*: expression of X causes expression of Y, or expression of Y causes expression of X . We may use additional knowledge to favour one direction over the other. For instance, if X is a known transcription factor while Y is not, we may favour the model X→Y (→ reads *causes)* over Y→X.

2. *Confounding*: a third, potentially unobserved, protein U causes expression of both X and Y, *i.e.* X←U→Y.
3. *Coincidence*: the correlation is solely by chance. If so, we shall observe diminishing correlations as we collect more data. Given the data, we can perform statistical inferences, for instance permutation test and bootstrapping, to test how likely we observe the correlation by chance.
4. *Conspiracy*: the correlation is due to deliberate manipulation of data or the sampling process. We may, for instance, create a good correlation by removing data from all cells where two proteins are not correlated.

Figure 1 illustrates the point that both causation and confounding can generate correlation. While statistical tools are available to assess the probability of a coincidence, and conspiracy may be discovered by science, statistical models alone cannot tell causation and confounding apart. For that we need causal inference, which requires a causal model of how data is generated, and operates the model with statistical modelling.

For some applications, correlation is all we need. In the toy example above, if our goal is to predict the expression of Y given the expression of X in a similar yet unobserved cell from the same population, correlation suffices and there is little need to distinguish between causation and confounding. Off-the-shelf statistical and machine-learning models are good enough, because they exploit correlations between input variables and the target variable.

Delineating causation from correlation is essential for many other applications. If our task is to predict the expression of Y in a cell where we inhibit the expression of X (Figure 1G), for instance by gene silencing, it is essential to know which of the models below is the correct one: X→Y, or Y→X, or X←U→Y. Because if X→Y is true, expression of Y shall become residual; given the other two models, expression of Y shall not be affected. If we do not have a model, collecting more data does not help us with the task.

The simple example illustrates three critical points: (1) causality is required to predict the outcome of intervention, (2) data alone is not able to tell causality, and (3) we need both scientific models, either prior knowledge or hypothesis, and data to infer causality.

The take-away message is that while statistical and machine-learning models are useful if we ask for correlation, we need additional tools to recover causality from correlations if our goal is to intervene in the system and predict the outcome. While correlation helps us to predict or even to explain[11,12], causality helps us to understand.

## 2.3 Causal modelling with Directed Acyclic Graphs (DAGs)

Directed Acyclic Graphs, or DAGs, are computational models of causal inference. DAGs contain nodes, which represent variables, and edges, which represent causal relationships. Like the two-variable cases described above, DAGs represent either knowledge or hypotheses about causal relationships between the variables. The edge, *i.e.* the causal relationship, can have any discrete, linear, or nonlinear functional form. A lack of edge between any pair of nodes means that we exclude the possibility that they have direct causality on each other. An indirect causal relationship may still exist in such cases due to the propagation of causality through the DAG, as we shall see below.

Three 3-node structures are prevalent in DAGs. Understanding these common structures, known as *motifs*, allows us to interpret more complex causal models.

1. The *pipe* (Figure 2A) describes the simplest chain of causality: X→Z→Y.
2. The *fork* (Figure 2B) describes a variable Z causing both X and Y: X←Z→Y.
3. The *collider* (Figure 2C) describes the situation where a variable Z is caused by both X and Y: X→Z←Y.

A pipe transduces causality in a chain, for instance *Vemurafenib binding to V600E-mutated BRAF→reduced BRAF signalling→reduced tumour size.* If the intermediate cause is manipulated, for instance by mutations enhancing BRAF signalling, the causal effect from drug on tumour size shall diminish or even reverse. Statistically speaking, X and Y are marginally correlated, and they are independent conditional on Z.

A fork, like the temperature example raised before, generates correlations between variables that do not cause each other. Example: *PI3K phosphorylation←RAS signalling→BRAF proliferation*. If the common cause is present, its both effects are correlated; if the common cause is absent, or if we focus on the subset of the data where the common cause takes the same value, the effects become independent. Similar to the situation of pipes, X and Y are marginally correlated, and they become independent conditional on Z.

Colliders behave very differently from pipes and forks. For instance, cells that lose the expression of the BOP1 (Block of Proliferation 1) gene become resistant against Vemurafenib[13]. The mechanism can be described by a collider structure: *BRAF V600E→increased phosphorylation of ERK1/2←loss of BOP1*, namely both BRAF V600E mutation (V600E for short) and loss of BOP1 (BOP1↓) increases ERK1/2 phosphorylation (↑pERK1/2). As long as either condition is satisfied, we can observe ↑pERK1/2. Assuming that they are the only two factors influencing pERK1/2, the collider creates a dependency between V600E and BOP1↓. When ↑pERK1/2 is true, either V600E or BOP1↓ or both must exist; otherwise, neither V600E nor BOP1↓ is possible. Even if V600E and BOP1↓ are statistically independent from each other, *i.e.* knowing V600E does not give us any information about whether BOP1 is lost or not, the two events become correlated once we observe ↑pERK1. In contrast to pipes and forks, X and Y are marginally independent but become correlated conditional on Z.

Knowledge of causal structure allows us to better recognize cognitive bias. We demonstrate this point with the example of colliders. With simulated data generated by the three 3-node structures in Figure 1, we show that we can remove or create correlation between *x* and *y* if we stratify by the third variable *z*, depending on the graph structure. The conclusion holds if *z* is a continuous variable and if we regress it out in a statistical model. Imagine now we face a dataset with two variables, *x* and *y,* we observe a good linear correlation, and our goal is to predict the value of *y* given a new set of values *x*. Does a linear model suffice? Not necessarily, because if there is a variable *z* forming a collider with *x and* y*,* and its value changes for the new set of values *x*, then our linear model is likely to fail. Even the simplest causal graphs show us it is important to know which factors affect the generation of data.

Besides the pipes, forks, and colliders, we highlight two 4-node motifs:

1. The *descendant* describes a variant of the *collider*, where the variable caused by two variables X and Y is not directly observed, but its *descendant,* namely another variable that is caused by the unobserved variable, is observed. Since we often use the variable *U* to indicate an unobserved variable, the DAG can be represented as: X→U←Y and U→Z. The structure, for instance, is useful to model the effect of known risk factors (X) and treatment (Y) on disease status (U). Though U may not be directly observed, a biomarker (Z), which is partially affected by U, can be measured.
2. The *paw* or *3-pan* describes a class of motifs, including the one shown in Figure 5A. The tail variable is known as *instrumental variable*. Models with instrumental variables are applicable in a wide variety of contexts, including Mendelian Randomization, a technique for target identification to be described in Section 4, and the analysis of clinical-trial data with non-compliance, *i.e.* when some patients in the treatment arm do not take the medicine as prescribed[14].

DAGs consisting of these and other motifs[15] can represent causal relationships between any number of variables. We refer readers interested in reading and understanding DAGs better to several outstanding tutorials[16–18].

Since DAG structure specifies marginal and conditional correlations, it is possible to rank models by their plausibility given data, and to learn the strength of causal relationships. The two tasks are known as *causal identification*, or *causal discovery*, in which the truth value of a claim of the form 'C causes E' is determined; and quantitative *causal estimation*, in which a numerical value *s* (strength) is estimated for a claim of the form 'C has an effect on E'. Causal discovery is more challenging than causal estimation, and discovered causal effects usually require causal estimation before they can be used for high-stake applications[19,20]. Nevertheless, it plays an important role in reconstructing causal networks for disease understanding and target identification, among others because we lack much biological knowledge in order to hypothesize how the observed data are generated. See Section 4 for details.

## 2.4 Relationship between causal inference and established modelling techniques

Causal inference completes statistical modelling and mechanistic modelling, two commonly used techniques in drug discovery and development (Table 1). Statistical modelling, including pattern-recognition-based machine-learning models, aim at identifying prediction-relevant features, and functions that transform the features to approximate the target variable and thereby exploit correlations. The models may or may not have a causal interpretation. On the other hand, mechanistic modelling uses mathematical models of biological processes to describe and predict how components, information, or energy of a system evolve with time. Examples include pharmaco-kinetics and -dynamics (PK/PD) and Quantitative System Pharmacology (QSP) models based on ordinary or partial differential equations (ODEs/PDEs), agent-based models, and Hidden Markov Models. While mechanistic models usually have a causal interpretation and are powerful for predicting outcome of intervention, we often miss information of important factors that affect the system's behaviour.

The three modelling approaches complement and benefit from each other. Reciprocal benefits between statistical and causal models are apparent. By discovering and quantifying causal relationships between observed variables, we can render statistical models causal interpretations. By using advanced machine-learning approaches, we may learn highly complex functional forms describing the causal effect[21], or discover causality from high-dimensional data[22]. Practically, since we are often limited by the volume of high-quality data, causal inference may likely profit from parsimonious statistical models and Bayesian approaches[23].

Results of causal inference help refine and improve existing mechanistic models by including causal factors and removing confounding factors. In turn, outputs from mechanistic models complement the current causal inference regime by predicting time-dependent outcomes of intervention, in particular those of components that regulate each other.

## 2.5 Causal inference for experimental data and observational data

The power of causal inference to connect and enhance existing modelling approaches is particularly prominent in the analysis of observational studies. Classically, we distinguish observational studies from controlled experimental studies. In a controlled experiment, we assign test objects, for instance animals in preclinical experiments, to groups. One group receives the treatment while the other group does not. If we go one step further to require that the grouping is randomized with regard to any relevant attributes of the test objects (passage, sex, body weight, etc.), then we have a randomised controlled experiment, a gold standard to establish causality. In an observational study, in contrast, we measure or survey members of a sample without trying to affect them. Such data can come from epidemiological studies, electronic health records (EHRs), insurance claims, and other data that come from natural (instead of controlled) experiments, for instance omics and behavioural data of healthy individuals and patients.

Several reasons make it imperative to use causal models to analyse data generated by observational studies. First, it allows us to integrate knowledge and hypotheses about inevitable biases in the data generation process. Second, we can investigate the causal effect of independent variables of interest while considering other variables that affect the outcome, known as covariates. Third, we can potentially resolve the effects of variables that influence both the independent variable and the outcome, known as confounding variables. If we analyse observational data as if we were handling randomised experimental data, without considering the causal structure underlying the data generation and collection, we may derive false, sometimes ridiculous, conclusions[24].

Even if a study is set up as a classical randomised experiment, various reasons may break the randomization and demand causal inference to identify the treatment effect. For instance, in clinical trials, noncompliance is a common issue, where some patients do not take the treatment as prescribed. To identify the real treatment effect while acknowledging the noncompliance, it is necessary to employ a causal model, with the *instrumental variable* model illustrated above as the simplest example. Other issues that have been successfully addressed by causal inference include missing data[25,26], for instance when patients drop out of the trial, and intercurrent events, *i.e.* events occurring after randomization which can either preclude observation of the outcome of interest or affect its interpretation (ICH E9(E1)), for instance development of anti-drug antibodies (ADAs)[27–30]. The book by Guido Imbens and Donald Rubin provides an outstanding introduction to the analysis of these and other aberrations from classical randomised experiments with causal inference[14].

Causal inference can leverage both experimental and observational data to generate insight. Eichler *et al.* proposed a new framework, known as *threshold-crossing*, that generates evidence by synthesising historical randomised clinical trials and real-world data resources[31]. A recent landscape assessment and elicited comments have also shown that causal inference penetrates and contributes significantly to clinical-trial study design and analysis by judging the suitability and integrating real-world data from observational studies[32,33].

Finally, we emphasise the importance of study design for both experimental and observational studies. It is widely acknowledged that well-designed randomised controlled experiments, both in preclinical drug discovery[34] and in clinical drug development[35], are the gold standard of verifying causal relationships. The quality and strength of evidence provided by an observational study and causal inference is also determined largely by the study design[36]. Study design should be scrutinised prior to causal inference in order to gauge whether we can answer the question of interest.

# 3. A step-by-step guide of causal inference

We introduce causal inference as an iterative process of six steps: *modelling → identification → estimation → refutation → refinement → application*. They are both illustrated in Figure 3 and detailed below. To assist practitioners acquiring hand-on experience with causal inference, we offer complementary interactive tutorials with case studies as Supplementary Information. The tutorials, implemented in programming languages Python (with *Jupyter Notebooks*) and R (with *Rmarkdown*), are available at
https://github.com/Accio/causal_drug_discovery.

### Step 1: Modelling

To start, we construct a DAG to model causal mechanisms that generated the observed data by synthesising common sense, scientific knowledge, or explicit assumptions. If multiple hypotheses exist, we may build several DAGs and subject them to analysis and comparison.

In case there is little knowledge and hypothesis available, we may apply causal discovery techniques to hypothesize about how data may be generated. Commonly used methods include Bayesian networks[37], factor analysis[38], encoder-decoder and other deep generative models[39]. Limitations of such methods include (1) they are often technically challenging (*NP-hard* problems), (2) the results often include many alternative hypotheses that explain the data equally well, and (3) the proposed models may not be causal, *i.e.* fail to be validated by interventional studies. Nevertheless, the discovered causal graphs, combined with expert's review and curation, may serve as reasonable starting points.

## Step 2: Identification

Once a causal model is set up, we test whether we can answer the question of interest quantitatively. The quantity that addresses the causal question, for instance *how strong is the effect of the drug on disease progression*, is known as the *estimand*. In this step our goal is to assess whether we can estimate the estimand from the data at all, because some graph structures prohibit us from doing so.

The concept of *estimand* is closely associated with *estimator* and *estimate*, all of which are core to statistical inference from data. To illustrate them with an example, assume that we give six randomly chosen patients drug treatment and make the following observations on disease progression within a predefined period: 1,1,0,1,1,0 (1=slower progression than placebo, 0= similar progression as placebo). Our goal is to estimate the probability *p* that the treatment slows down disease progression.

1. The *estimand* is the quantity of interest, in this case the probability *p*. Apparently the estimand is defined by the objective of the experiment and data analysis.
2. The *estimator* is the rule for estimating the estimand. Multiple estimators may exist for the same estimand. In the given example, the maximum-likelihood estimator of *p* is $\hat{p} = \frac{x}{n}$, and a Bayesian estimator with a non-informative prior is $\hat{p'} = \frac{x+1}{n+1}$.
3. The *estimate* is the numeric value that we get by analysing the data by following the rule of the estimator. In this case, $\hat{p}$=0.67 and $\hat{p'}$=0.71.

In a causal model, the estimand is usually the causal effect pointing from a treatment (cause) to its target (effect) variable, which can be a regression coefficient or other numeric values that quantify the strength of the causal relationship. The causal graph model structure determines which estimators are available, and the numerical values of estimates are gauged by the users to interpret the results and challenged by refutation analysis.

An interesting and important result from theoretical studies is that we can assert whether an estimand can be identified using graphical models alone, independent of the choice of the estimator and without access to the data. This means that we can tell whether the question can be answered at all by causal inference *before* we collect or generate any data, which is usually the most time- and labour-consuming step in knowledge acquisition.

In this step, we need to identify the target estimand by specifying the quantity of interest in the DAG model. Usually we use software to identify whether the estimand can be estimated based on the graph structure. If the estimand is not identifiable, we have little choice than modifying our model, redefining our question, or admitting that the problem cannot be solved with causal inference. Otherwise, the model is said to be *identifiable*, and we can continue with the estimation step.

A causal model is intrinsically a generative model, namely it can be used to generate simulated data. Once the identifiability is established, a common practice is to simulate data with the model with specified parameters, to run the inference with simulated data, and to examine both whether the simulated data reassemble observed data, and whether the estimate is consistent with the input parameters. Inference with simulated data offers us experience with the model, opportunities to identify bugs and flaws of the model before the real run, and confidence in the model and the choice of estimator.

## Step 3: Estimation

Next we collect existing or generate new data according to the DAG. The importance of quality control of data cannot be overestimated: *garbage in, garbage out*. Data may be filtered so that only the subset that satisfies the DAG structure is used for the estimation.

Once we have decided on the estimand and the data, we can estimate the causal effect by identifying an appropriate estimator and deriving the estimate from the data, which is usually done by software (see Box 2 for popular open-source software packages for causal inference).

The choice of estimator depends on the DAG structure and by the nature of data. The popular *DoWhy* package, for instance, offers diverse estimators depending on the graph structure, including (1) (generalised) linear regression, (2) stratification, matching, or weighting by propensity score, namely the probability of each unit being assigned to a treatment group given a set of observed covariates, (3) using instrumental variables, and (4) machine-learning based estimators implemented in the *EconML* package. While the choice of appropriate estimators is context- and question-specific, several measures can help us make better choices, especially (1) using simulated data, (2) learning from case studies and past experience, and (3) applying refutation techniques discussed in the next step.

## Step 4: Refutation

While estimation is commonly deemed as the most important step in causal inference, in high-stakes applications such as drug discovery and development, we need to test the robustness of our estimates. This is achieved by *refutation*, a collective name of many modelling techniques to test the strength, validity, and our confidence of the estimated causal effect.

Commonly used refutation techniques include

- *Sufficiency test with unobserved random common cause*: Does the estimate change when we add an independent random variable as a common cause (confounder) to both the treatment and the outcome? The estimate should not be too sensitive.
- *Placebo treatment*: What happens to the estimate when we replace the true treatment variable with an independent random variable? The effect should go to zero.
- *Dummy outcome*: What happens to the estimate when we replace the true outcome variable with an independent random variable? The effect, again, should go to zero.

In addition, one can perform data partition or bootstrapping to estimate the variability of the estimates. Furthermore, bespoke refutation techniques can be used to address application-specific questions, for instance non-compliance in randomised clinical trials[40]. We refer interested readers to a recent review about these and other techniques known as sensitivity analysis[41].

Statistical causal inference is not the only way to identify causality (Box 1). Alternative evidence from other models, for instance, results from mechanistic models and results from randomised controlled experiments, should be used to strengthen or challenge our belief in the output of the causal model.

### Step 5: Refinement

Even when the estimated causal effects withstand refutation analysis, they are seldom the sole goal of our investigation. Given that we lack a comprehensive understanding for virtually all biological problems, we often need to *refine* the model, which includes selection from multiple models, addition of new variables, interpretation of latent variables and of estimates, and answering *what-if* questions. The refined model can be used on one hand to guide design of new experiments to further refute or update the model, and on the other hand to guide interventions.

### Step 6: Application

One of the ultimate goals of causal inference is to perform interventions in the real world. The outcomes can be further analysed in the causal inference framework, therefore closing the loop.

In contrast to a 'hypothesis-free' paradigm, where one expects to learn mechanisms generating the data from data alone, the cycle of *modelling, identification, estimation, refutation, refinement,* and *application* integrates knowledge, hypothesis, and data to address scientific questions. Interdisciplinary teamwork is required to construct explicit models of causality, to collect estimand-specific data, and to identify causal relationships.

# 4. Literature review and case studies

## 4.1 Literature review

To gain an overview of applications of causal inference in drug discovery and development, we compiled a list of publications by querying the MEDLINE/PubMed database. We found more than 800 scientific publications (Supplementary Table 1). As comparison, we also queried publications on machine learning and artificial intelligence and stratified the number of publications by years (Figure 4A). We note two interesting patterns: (1) publications on causal inference in drug discovery have increased almost steadily since 1990, and (2) the topic receives much less coverage compared with machine learning and artificial intelligence. Despite that the scope of the latter two concepts is admittedly much broader, the patterns underscore the importance of populating the causal mindset and language among practitioners.

We classified the literature by applications, and observed that causal inference has been applied throughout the value chain of drug discovery and development (Figure 4B and 4C). The majority of publications was about analysis of observational studies, especially drug safety, pharmacovigilance, and real-world data. Methodological papers and publication reporting applications for target identification and assessment, clinical trial design and analysis, and biomarker studies followed with distance. Applications in other areas, in particular preclinical and translational research, are still scarce.

Given the broad application of causal inference, and given that applications for clinical data, real-world data, and observational studies have been extensively reviewed elsewhere[32,33,42,43], we focus below on case studies of causal inference for disease understanding and target identification in translational research.

Translational research includes activities that establish the causal relationship between drug candidates and disease progression. It has two interlinked components: forward and reverse translation. Forward translation predicts drug's dosing and risk-benefit ratio in patients with *in silico* models and data from *in vitro*, *ex vivo*, *in vivo* animal models, and microdosing human studies. Reverse translation informs and improves decision making in forward translation by analysing data collected in clinical trials, real-world data, as well as data generated with clinical-stage or marketed molecules in preclinical models. Forward and reverse translation form a closed loop and complement each other to validate or refute the proposed causal relationship between treatment and patient's health status.

While translational research has many facets, we choose disease understanding and target identification to highlight the impact of causal inference for two main reasons. First, leveraging the causal engine in the target identification and selection phase may have the strongest potential in reducing costs of disease and of drug discovery to society[44–46]. Second, applications in this area involve heterogeneous and high-dimensional data, for instance EHRs and genomics and other omics data. Though the data type varies in other tasks of translational research, for instance risk-benefit assessment, biomarker selection, and patient stratification and enrichment, the concepts, software tools, and practice can be transferred.

## 4.2 Learning causal associations from natural experiments and observational studies

While target identification was driven by druggability and *in vitro* or animal disease models for a long time, reverse translation is upending this pattern. Causal inference contributes evidence to support or disaffirm drug targets by analysing data from natural experiments and observational studies. The need for causal inference is particularly strong in emerging fields such as microbiome medicine where high-quality data are relatively scarce and associative analyses are still prevalent[47]. Similarly, causal inference is needed when the study of disease aetiology is challenging, for instance in neuroscience. Path analysis, a precursor to and variant of causal inference, has been applied to prioritise or refute targets in neurogenesis, Parkinson's Disease, and autism spectrum disorder[48–50]. Recently, researchers used path analysis to dissect the quantitative contribution of Alzheimer's biomarkers and risk factors to cognitive impairment and decline by analysing longitudinal biomarker data[51]. The study reported that about 16% of variance in both cross-sectional and longitudinal cognitive impairment was accounted for by amyloid-beta and about 46% by tau. The findings suggest that both removing Aβ and completely shutting down Tau downstream events may only be partially effective in slowing down cognitive decline or reversing cognitive impairment. The results provide predictions of the theoretical boundary of efficacy of existing and new molecules modulating these targets. By integrating insights from such studies and from other modelling approaches, for instance the Q-ATN model developed by Norman Mazer and colleagues[52], we may be able to better understand disease aetiology and to propose better targets based on real-world evidence.

Application of causal inference on multi-omics (genomics, transcriptomics, proteomics, metabolomics) data leads us to identify causal genes, pathways and gene regulatory networks that have a direct effect on disease states.[53–56] Genome-wide association studies (GWAS) have mapped the genetic architecture of common diseases in humans[57], and have identified genetic loci that affect drug response and susceptibility to adverse drug reactions[58]. Drugs targeting genetically supported targets tend to be twice as likely to be successful as other drugs.[59,60] Exploiting GWAS data in drug discovery is, however, challenging because there are generally hundreds to thousands of genetic risk variants, mostly lying in non-coding genomic regions and each typically contributing only a small amount of risk.

"Mendelian randomization" (MR)[61,62] is a statistical approach that uses genetic variants as randomised instruments ("natural experiments") to identify causal associations between heritable traits. It is based on the fact that genotypes are independently assorted and randomly distributed in a population by Mendel's laws, or in a less strict manner, independent of each other outside haplotype blocks,[63] and not affected by environmental or genetic confounders that affect both the exposure (target manipulation) and the outcome (disease state). If it can be assumed that a genetic locus affects the outcome only through the exposure, then the causal effect of the exposure on the outcome can be derived from their relative associations to the genetic locus, which acts as an instrumental variable[64] (Figure 5A).

By integrating GWAS with studies that map genetic effects on the transcriptome or proteome (expression or protein quantitative trait loci; eQTLs or pQTLs, respectively), MR can identify causal associations between molecular traits and disease. When a molecular QTL (mQTL) and GWAS locus overlap, MR suggests causal variants that affect phenotype via regulating gene expression, and can distinguish them from pleiotropic causal variants that affect gene expression and phenotype simultaneously, or multiple causal variants that are in linkage and regulate expression and phenotypes independently[65]. When applied to proteomic data, inferred causal associations between proteins and disease can suggest drug repurposing opportunities[66] or new candidate drug targets[67,68]. An advantage of MR is that it can be performed using summary statistics alone[69]. A limitation is that MR tests effects of molecular traits on diseases one-by-one and that no molecular pathways are reconstructed. This is important because the candidate causal factors may act indirectly or redundantly with other factors, or the candidate factors may not be druggable themselves but affect intermediate druggable targets.

Statistical model selection can orient the direction of causality among correlated genes or proteins by merging genetics and functional genomics data from individuals in a segregating population. This insight was first put forward in a visionary paper by Jansen and Nap[70], and then formalised in a Bayesian model selection procedure[71]. This procedure compares the conditional independence relations implied by an independent, causative or reactive DAG among QTLs and pairs of genes or complex traits to infer the direction of causality from the model that best fits the available data (Figure 5B).

Further theoretical developments have extended this model selection procedure by quantifying the statistical significance of the best fitting model to allow multiple testing correction[72–74], introducing additional DAGs with hidden confounders in the model comparison[75,76], and providing an efficient software implementation that can handle the size of modern genome-wide studies[75]. Statistical significance is estimated by expressing the conditional independence implications of the tested DAGs as combinations of likelihood ratio tests. Either their maximum $P$-value is treated as an omnibus hypothesis test[73,74,76], or false discovery estimation techniques are used to express the results of the individual tests as probabilities of their null or alternative hypothesis being true, which can then be combined by the usual rules of probability theory[72,75]. Accounting for hidden confounders (e.g., coregulation by an unknown upstream factor) requires additional assumptions on the genetic instrument, usually satisfied if the instrument is a *cis*-acting eQTL or pQTL for the "exposure" variable, and involves a trade-off between a high false negative rate or an increased false positive rate, when hidden confounders are ignored or included in the model comparison, respectively[75,77].

Bayesian networks combine the results of pairwise causal inferences into a comprehensive causal network model of the underlying biological system. Bayesian networks are probabilistic machine learning models for expressing both prior knowledge and inferred conditional independence and causal relationships among variables[78–80]. A Bayesian gene network consists of a directed graph without cycles, which connects regulatory genes to their targets[81,82] (Figure 5C). The structure of a Bayesian network is inferred from the data using score-based or constraint-based methods[79]. Score-based methods maximise the likelihood of the model, or sample from the posterior distribution using Markov chain Monte Carlo (MCMC), using edge additions, deletions or inversions to search the space of DAG structures. Score-based methods have been shown to perform well for reconstructing Bayesian gene networks from genetics and genomics data using simulated and model organism data[83–86]. Constraint-based methods first learn the undirected skeleton of the network using repeated conditional independence tests, and then assign edge directions by resolving directional constraints (v-structures and acyclicity) on the skeleton. They have been used for instance in the joint genetic mapping of multiple complex traits[87].

When individual-level genotype and gene or protein expression data are available from the same individuals, the results of causal inference tests are used to constrain the DAG search space[81–87]. This is done most elegantly by expressing the joint likelihood of observing both data types as a standard likelihood term for observing the expression data given the DAG and an additional likelihood term for observing the genotype data given the expression data and causal interaction implied by the DAG[81,82,85–87] (Figure 5C). For large systems, the latter term can be expressed as an independent product over pairwise causal edge probabilities inferred from the model selection procedure (Figure 5B)[86], whereas for sufficiently small systems, genetic association mapping and Bayesian network reconstruction can be performed simultaneously[81,82,85,87].

Using causal inference and Bayesian networks to identify causal genes, pathways and gene regulatory networks from multi-omics data has resulted in a system-level understanding of complex diseases and proposed novel target genes. Bayesian networks were used to map the genetic architecture of human liver gene expression and propose causal genes at loci associated with risk for type 1 diabetes, risk for coronary artery disease, and plasma low-density lipoprotein cholesterol levels[88]. Using genomic and transcriptomic data from post-mortem samples from four brain regions of late-onset Alzheimer disease cases and nondemented individuals, the gene *TYROBP* was identified as a key regulator of a causal immune and microglia gene network[89]. A role for TYROBP in the morphology of amyloid deposits and level of TAU phosphorylation was later confirmed experimentally.[90–92] Another analysis of genomic, transcriptomic, and proteomic data from four brain regions from 315 AD cases identified *VGF* as a key regulator in multiple AD causal networks, and overexpression of VGF in an Alzheimer mouse model was shown to reduce neuropathology.[93] Using genomic and transcriptomic data from seven vascular and metabolic tissues from coronary artery disease (CAD) cases undergoing surgical intervention, CAD-causal Bayesian gene networks were identified that replicated in data from matching tissues in the Hybrid Mouse Diversity Panel[94]. It was later found that one of these causal networks is susceptible to treatment by antiretroviral therapy drugs, and that targeting it may help reduce CAD side effects of ART drugs[95]. Further computational analysis of multi-omics causal networks in the context of protein interaction and pharmacological databases can identify established and novel druggable targets and target tissues[96].

To bridge from the molecular scale to disease states, clinical phenotypes are not usually modelled as nodes in the causal network. Instead genome-wide omics data is first partitioned into co-expression modules associated with clinical traits[89,93,94,97], causal networks are learned for each module separately, and connected in a higher-level Bayesian or coexpression network where each node is a module "eigengene", a representation of the module's expression profile by its first principal component[98] (Figure 5E). This higher-level network can model tissue-specific and inter-tissue communication processes that are the result of the "collective" states of molecular networks and that are more proximal to the physiological disease processes[55,94,97,99].

As in almost all other areas of genomics, single-cell technologies are creating new opportunities for learning causal disease mechanisms from population-based data[100]. Cost-effective strategies for generating single-cell RNA-seq data across individuals have been developed[101,102] and the first large-scale population genetics studies have been published recently[103–107]. A particularly attractive application will be to exploit the fact that allele-specific expression can be mapped in single cells[108]. This will allow us to quantify eQTL effect sizes[109] at the level of an individual instead of population, which will lead to a better understanding of the variation of causal effects across individuals. Moreover, since allele-specific expression can also be linked to transcriptional bursting[108], an intrinsically random event, it can potentially be used as another type of randomised instrument or natural experiment to infer causal associations from observational single-cell RNA-seq data.

## 4.3 Learning causal associations from controlled perturbation experiments

Having considered the impact of causal inference on reverse translation, now we turn to its impact on forward translation by analysing controlled perturbation experiments (Figure 5D).

For target-based discovery programs, canonical gene knockout or perturbation experiments in model systems establish causality in a manner similar to randomised controlled trials. When these experiments are combined with genome-wide readouts, they provide insights into the direct and indirect causal targets of the perturbed gene. Systematic small molecule and perturbation screens using RNAi technology followed by reduced representation transcriptome sequencing have been performed in cancer cell lines[110]. CRISPR-based technology combined with single-cell RNA sequencing is now rapidly expanding the scope of conducting genome-wide perturbation screens in human cells[111–115].

Causal inference from perturbation experiments allows us to disentangle direct from indirect effects and predict the outcome of a new intervention or perturbation. Nested effect models are probabilistic graphical models to infer a genetic hierarchy from the nested structure of observed perturbation effects[116,117]. A more formal approach uses causal DAGs and *do*-calculus to model causal effects from perturbation screens[118,119].

Pioneers and experts in phenotypic drug discovery have long realised the importance of establishing causal relationships. Moffat *et al.*[120] proposed a model of *chain of translatability*, namely from the chemical matter to the assay phenotype, to the preclinical disease model, and finally to the human disease. We note the multiscale nature of this chain of translatability and the resemblance of challenges faced by target-based and those by phenotypic programs except for the target identification and assessment question. A key open question is whether causal inference, and in particular causal discovery techniques, can help phenotypic programs identify molecular targets and mode of action.

A fundamental issue with forward translation is that cellular or animal models differ from human physiology. Therefore a critical additional request is that the identified link must be persistent in humans, in the form of *in vitro-in vivo* translation, governed by physiological constraints, and between-species translation, governed by evolutionary forces.

In order to decide whether experimental findings can be extrapolated across domains that differ both in their distributions and in their inherent causal characteristics, we need prior scientific knowledge about the invariance of certain mechanisms is available and represented in "selection diagrams", DAGs in which the causal mechanisms are explicitly encoded and in which differences in populations are represented as local modifications of those mechanisms[121,122]. Switching between populations corresponds to conditioning on different values of the variables that locate the mechanisms where structural discrepancies between the two populations are suspected to take place. The transportability problem can then be formulated and decided using *do*-calculus and graphical criteria on the selection diagrams.

Translational study is undergoing a profound change. While animal models are traditionally used as a bridge between *in vitro* experiments and studies with human, the field is witnessing strong interest and investment in human model systems such as stem-cell-derived cell lines, primary cells, organoids, organ-on-a-chip, or other microphysiological systems. They are currently being tested for many purposes, especially efficacy and safety assessment, prediction and modelling of pharmacokinetic and pharmacodynamic parameters, biomarker identification, and disease modelling[123].

From the point of view of causal inference, the key question is how much the causal model relevant for the disease and the drug candidate in human systems is conserved in model systems, let it be *in vitro*, *in vivo*, or human model systems. To answer this question, it is necessary to establish causal models with each model system, and to compare the models between the systems. So far, the translational value of most human model systems is judged by *looking alike*, i.e. to assess similarity in morphology and/or *omics* profiles with regard to primary human material. In order to mimic the causal relationship, more work is required to assess these systems by *functioning alike*, i.e. to assess their response upon perturbation and conditions that mimic disease aetiology and development[124,125].

# 5. Discussions and conclusions

This review addresses three questions: when do we need a causal model, how to perform causal inference, and which activities in drug discovery have been or will be empowered by causal inference. We argue that causal models are essential for predictive modelling of outcome of intervention, for answering *what-if* questions, and for understanding disease biology and why drugs work (or not). We introduced causal inference as a six-step iterative process with hand-on examples in the complementary tutorial. And we performed literature review and discussed recent application of causal inference in translational research.

As discussed in the introduction, causality is multiscale. This renders causal inference a suited tool to investigate interactions between drugs and human biological systems, which are *per se* multiscale[126]. By leveraging advances in statistical modelling and mechanism-based multiscale modelling, it has the potential to transform heterogeneous data into unified knowledge, which helps us understand the impact of genetic, epigenetic, and environmental factors on drug action[127].

An emerging consensus is that the quality of *in silico* model predictions outweighs the speed of prediction, and qualitative statements that enable scientists to focus on promising targets and regions of chemical space are often more impactful than quantitative predictions for decision making[128,129]. In line this, we foresee three key opportunities for causal inference: (1) high-quality prediction for predictive modelling of interventional outcomes, including out-of-distribution predictions[128,129], (2) individualised causal inference and estimation of heterogeneous treatment effects[130], and (3) causal- and counterfactual-based decision making[16,131].

Besides the opportunities, we foresee three major challenges to be overcome by researchers and practitioners of causal inference in the coming decade: (1) the methodology of causal inference needs to be further developed to embrace the reality and complexity of biology, (2) the community needs to share and have open access to high-quality data and models, and (3) we need a change in both language and mindset.

The methodology of causal inference warrants further research. DAGs, though very powerful, have intrinsic limitations when being used to model biological systems. Edges in DAGs must be acyclic, *i.e.* we are never trapped in a closed loop by following the directions of edges. DAGs are therefore not suited for analysing reciprocal causal relationships, though they are prevalent in biology, across scales from molecular interactions[132] to consciousness[133]. To model systems with such relationships, one can either take advantage of longitudinal data to identify causal relationships[134], or model the net output of nodes involved in reciprocal relationships as a variable instead of modelling individual nodes, or employ computational models other than DAGs, for instance graph neural networks and other representation learning techniques[135]. Both the theory and the software required for performing such analyses need further research and development. Despite recent progress discussed above[22], it is still challenging to perform causal inference with high-dimensional data, for systems with multiple complex traits, *i.e.* many factors with small effect sizes[136–138], and for complex adaptive systems where feedback loops are abound[139–141].

Open contribution and access to improve biological models is essential for the community to refine and leverage causal models. Resources of causal models are emerging quickly[142], and community-wide efforts like DREAM challenges have been hosted to identify causal relationships from *omics* datasets[143]. Nevertheless, the causal language remains foreign to many researchers, we do not always distinguish correlation from causation explicitly, and in many cases scientific findings are not represented as causal models. While it remains a complex and unsolved problem of how to populate the causal language and how to encourage researchers to share data and causal models, we are cautiously optimistic about further development because of the potential gain. The Linus's Law formulated by Eric S. Raymond[144], *Given enough eyeballs, all bugs are shallow,* is a motivation for all of us to adopt the causal language and share the causal models: to share and to refine causal models in a community is to understand how biology and drugs work collectively.

Several reasons may make practitioners of drug discovery and development, including many colleagues we have interviewed, wary of causal inference. We need the front-up payment of bringing up knowledge and hypothesis, often exposing our naivety and lack of knowledge. In a real-world setting, setting up a model is particularly challenging because we often have incomplete information and knowledge about human biology, pharmacology and toxicology of drugs, not to mention the abundance of invisible or unquantifiable quantities and events that affect efficacy and safety of drugs. We have to give up the hope about 'learning from data alone' and 'hypothesis-free approaches'. We know that machine-learning models such as deep neural networks and graphical neural networks can learn patterns if they are trained with enough data. Isn't it better to accumulate more data and let the machine find out the causality? Why do we stick to and propose causal inference?

We welcome these questions and doubts. They lead us to believe that we need a change in both language and mindset to think and talk about causality. We argue that (1) causality exists beyond data and it is not possible to learn causality from data alone without prior knowledge and hypotheses, and (2) the central task of drug discovery and development is to discover and validate multiscale causal relationships. The true causal network, in contrast to causal DAG models, can contain feedback loops and therefore cannot be subject to the analysis introduced here as a whole. However, the network can be separated into smaller sub-networks, each verified by the six-step cycle of causal inference, and the causality propagates between the scales. The network connects the drug molecule with disease outcome via its interactions with biological molecules. The interactions emerge as pharmacology and toxicology on cellular, organ-and-system, and population levels. Furthermore, the network can be enriched with individual traits such as genetics, medical history, lifestyle, and environmental factors to allow population analysis.

In conclusion, causal inference offers a principled approach to model- and data-driven predictive modelling and decision making. It complements statistical and mechanistic models to empower an understandable synthesis and integration of knowledge and data across scales. We foresee that further development of its methodology, open-accessible causal models and high-quality data, and a broader adoption of the causal language and mindset will fuel both forward and reverse translation to seek and to prove causality with new drugs.

# Acknowledgements

Both authors thank colleagues, students, and team members for discussions, inputs, and feedback for the manuscript. JDZ wishes to explicitly thank input from the members of the Predictive Modelling and Data Analytics (PMDA) chapter, especially Zhiwen Jiang, Milad Adibi, Juliane Siebourg-Polster, Tony Kam-Thong, Balazs Banfai, Sarah Morillo Leonardo, Martin Ebeling, as well as visitors of the weekly *Bioinformatics Club*.

# Funding

TM is supported by grants from the Research Council of Norway (grant numbers 312045 and 331725). JDZ's work is funded by F. Hoffmann-La Roche Ltd.

# Conflict of interest

The authors declare no conflict of interest.

# Supplementary material

Accompanying this review, we offer a set of open-source, hand-on tutorials for causal inference at https://github.com/Accio/Causal-Inference-Python-R. The tutorials are written as interactive, reproducible notebooks in *Rmarkdown* and *Jupeter Notebook* formats. They illustrate concepts and practices of causal inference with both toy and real-world examples analysed with selected popular open-source tools (see Box 2).

Supplementary Table 1 offers the list of literature on causal inference in drug discovery and development that we reviewed in Section 4.1.

# Figures, Tables, Glossary, and Boxes

## Figure 1

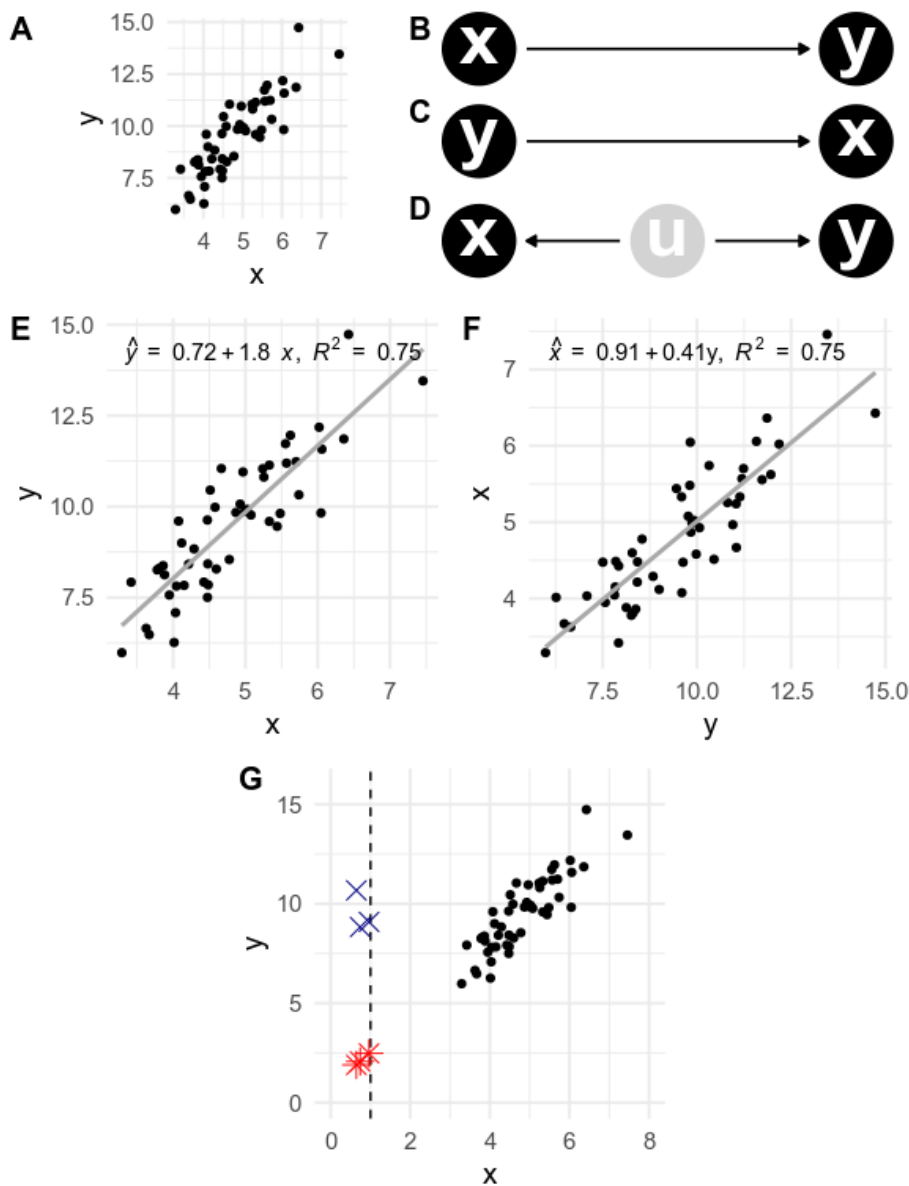

**Figure 1: Distinguishing between correlation and causation.** (A) Correlation between two variables *x* and *y*. The data were simulated by a linear relationship. (B-D) Three causal models that can generate the correlation observed in (A): two direct causal relationships (panel B and C) and one with a confounding variable (panel D). Coincidence and conspiracy as reasons generating correlations are not shown. (E-F) Data and statistical models alone

cannot tell the direction of causation, or whether confounding variables exist. Panel E shows the linear regression with *x* as independent variable and *y* as dependent variable using data shown in panel A. Panel F shows the regression with the same dataset, with the role of *x* and *y* swapped. The correlation coefficients ($R^2$) of both regression models are identical. (G) Predicting the outcome of interventions requires causal models. We use *x* and *y* to *denote* expression of two proteins X and Y in a population of cells. And now we reduce the expression of X artificially to 1.0 (dashed line in black). Depending on the causal structure, the outcome of Y may be near to the red stars (model B) or blue crosses (model C or D).

# Figure 2

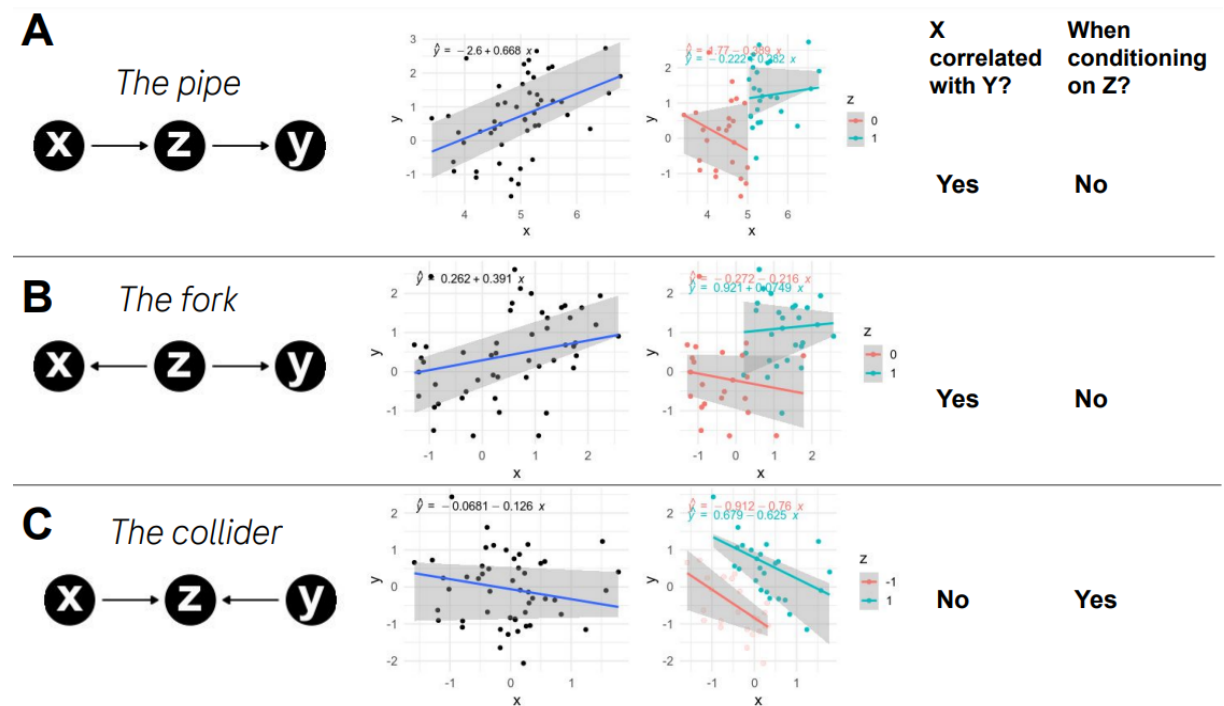

**Figure 2: : Common 3-node structures in DAGs**. (A) Left: the pipe structure consists of a chain of three variables. The model was used to generate a simulated dataset of 50 data points. We specified that *x* follows a Gaussian distribution with mean of 5 and standard deviation, *z* takes the value of 0 if *x*<5 and 1 otherwise, and *y* follows another Gaussian distribution with mean defined by 2*z and standard deviation. Middle: simulated data visualised with scatter plot. Each dot represents one data point. Both plots show *x* on the X-axis and *y* on the Y-axis. The positions of the points are the same in both plots. In the left plot, the regression line (blue) and its confidence interval (grey) are shown for y~x; in the right plot, the regression lines and confidence intervals are shown dependent on the value of Z. A simple visual assistance is that if *x* and *y* are marginally (mid-left plot) or conditionally (mid-right plot) correlated with each other, then the confidence interval should *not* contain any horizontal grid line. In this case, it is clear that while *x* is correlated with *y*, the correlation is broken if we condition on *z*. This summary is shown on the right panel of the plot. (B) Similar to A, with the fork structure on the left. Simulation rules (data points N=50): *z* follows a Bernoulli distribution with a probability of success of 0.5. Both *x* and *y* follow normal distribution with mean defined by *z* and standard deviation. The interpretation of middle and right panels is comparable to panel A. (C) Similar to A and B, with the collider structure on the left. Simulation rules (data points N=50): Both *x* and *y* follow normal distribution with 0

mean and standard deviation. The value of *z* is 1 if x+y>0 and -1 otherwise. The interpretation of middle and right panels is comparable to panel A and B. The code to generate both the simulated data and the visualisations is available at
https://github.com/Accio/causal_drug_discovery/blob/main/2021-12-CausalSalad.Rmd.

# Figure 3

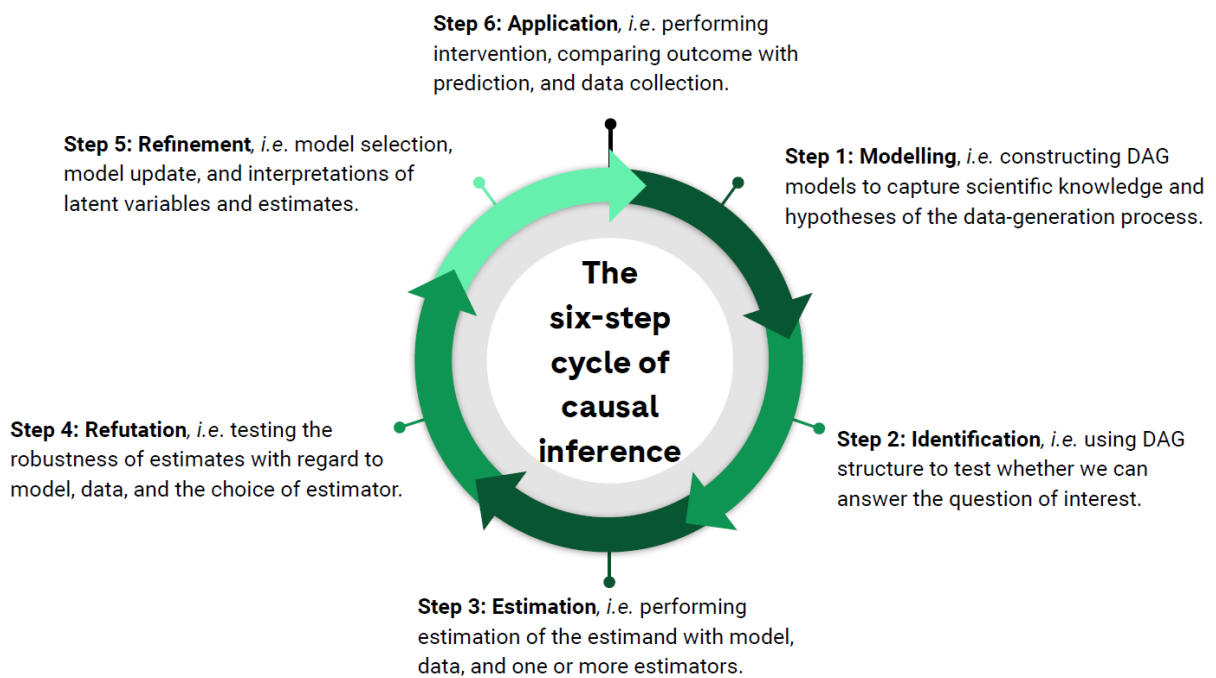

**Figure 3: A six-step model of causal inference, detailed in Section 3.**

# Figure 4

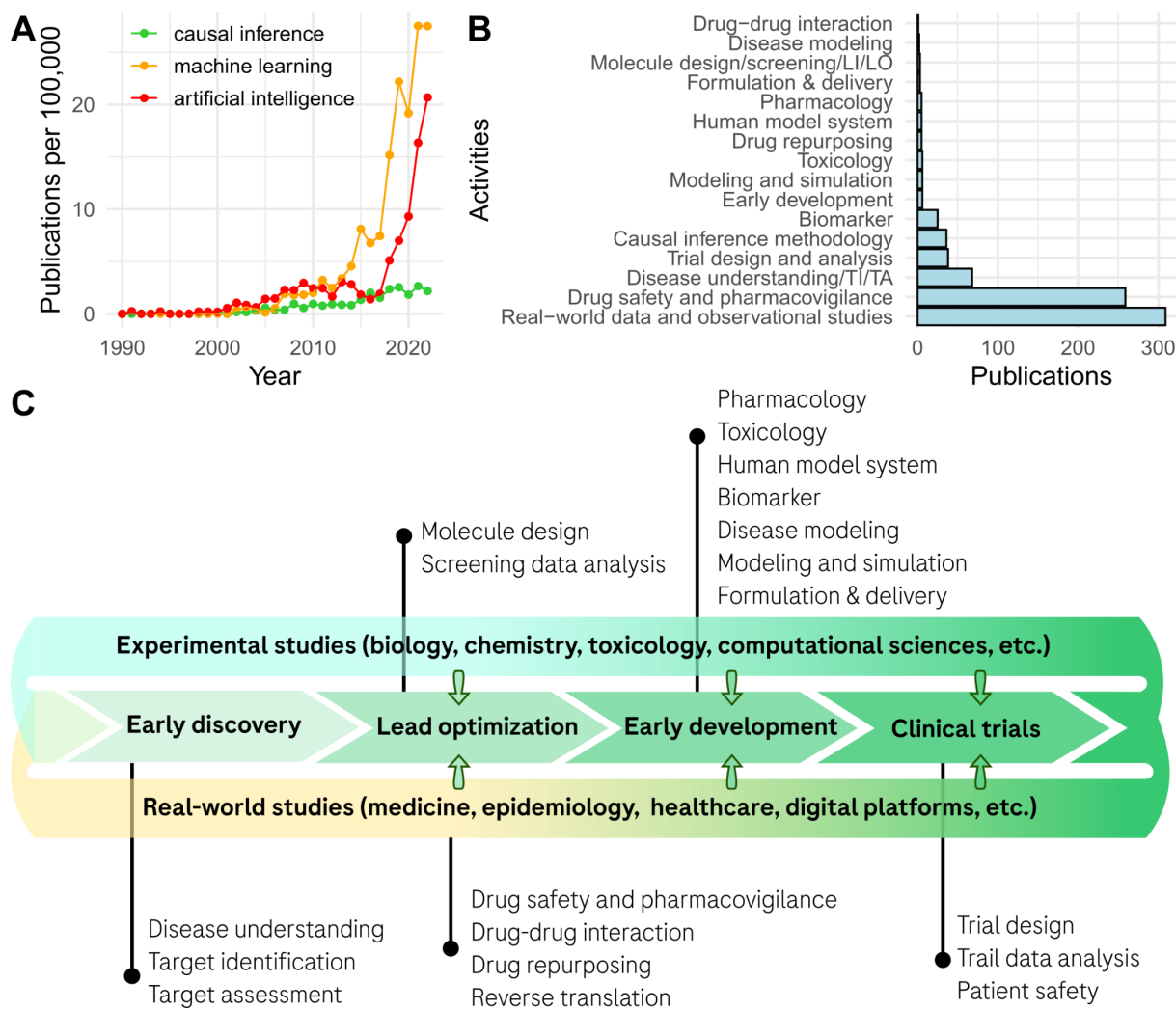

Figure 4: **Established and emerging applications of causal inference along the value chain of drug discovery and development.** (A) Publications indexed by MEDLINE/PubMed with keywords *causal inference* (green), *machine learning* (yellow), and *artificial intelligence* (red). Only peer-reviewed papers relevant for drug discovery and development are included. Numbers of publications are divided by the total number of publications and multiplied by 100,000. (B) Classification of publications on causal inference in drug discovery and development into activity categories. We read the abstract, and if available full text, of identified causal-inference publications and manually categorised them. The activities are reversely ordered by the number of total publications. (C) We positioned activities found in the literature review in the context of both forward and reverse translation. In the middle we show a simplified diagram of forward translation consisting of four steps: early discovery, lead optimization, early development, and clinical trials. Above and below we show reverse translation, i.e. analysis of experimental and observational real-world data, and its feedback to forward translation. Applications of causal inference are highlighted besides lollipops.

Figure 5

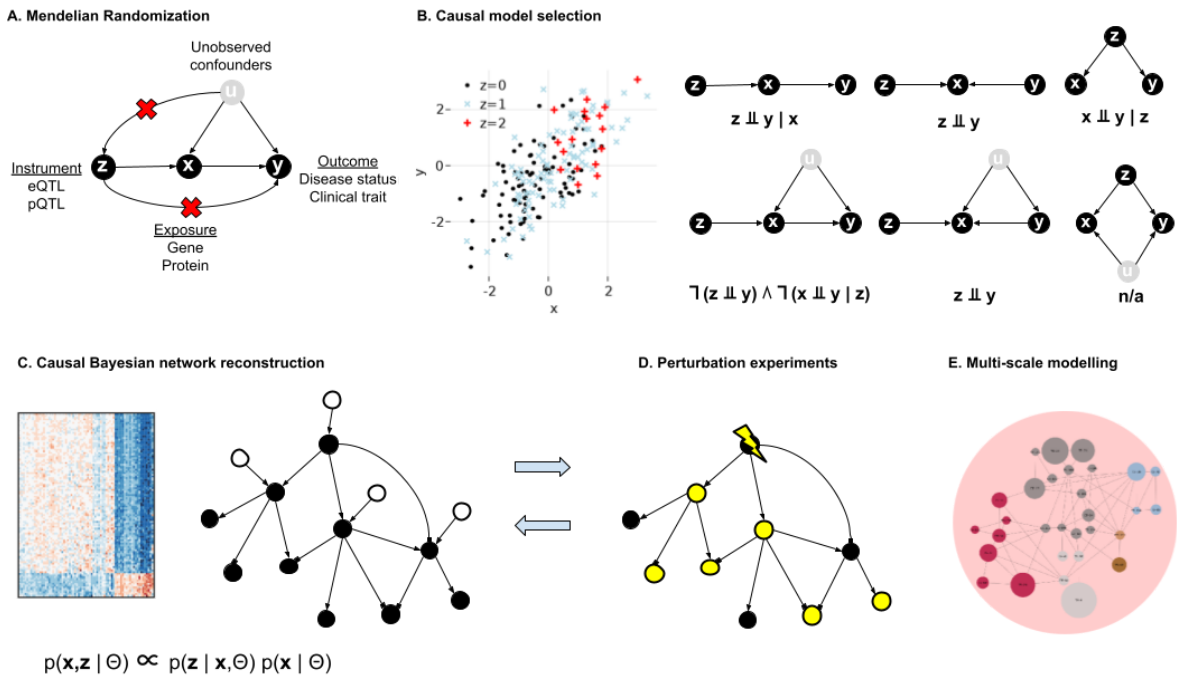

**Figure 5: Causal inference for disease understanding and target identification.** (A) Mendelian Randomization (MR) estimates the causal effect of an exposure ("x", e.g. a gene or protein expression level) on an outcome ("y", e.g. disease status or clinical trait) using a genetic instrument for the exposure ("z", an eQTL or pQTL for a gene or protein exposure). MR is applicable if the instrument is independent of unobserved confounders ("u") and if the outcome is associated with the instrument only through the exposure (no alternative paths) (indicated by red crosses). (B) Simulated scatter plot of two coexpressed genes X and Y, with samples colored according to the genotype of a genetic marker Z for X (e.g. a cis-eQTL for X). Model selection determines the causal model underlying the data by testing the conditional independencies implied by each causal DAG. Representative causal models without (top) and with (bottom) unobserved confounders are shown. Model selection assumes that the edge Z→X must be included in the model and that Z is independent of any unobserved confounders. (C) When a larger set of coexpressed genes is considered (left), the data is modelled by a causal Bayesian network (right). Black nodes represent genes (x) and white nodes genetic instruments (z) that are used to orient the causal directions of gene-gene edges, and the Bayesian network represents the probability distribution of jointly observing x and z given a set of model parameters Θ learned from the training data. (D) Controlled perturbation experiments give direct information about the variables causally downstream of the perturbed node (yellow nodes) and can be used to refute or refine networks reconstructed from observational data. (E) Models inferred at one scale (e.g. cell-type or tissue-specific gene or protein networks) can be integrated into higher-level models. In the figure, nodes represent tissue-specific causal Bayesian networks (node colour, tissue; node size, network size) and edges represent "eigengene" similarities. Panel E obtained from Talukdar, Husain A., et al. Cell systems 2.3 (2016): 196-208 under CC-BY-NC-ND licence.

## Table 1

A comparison of data modelling techniques, adapted from Table 1.1 from Peters (2017) [140]

|  | Examples | Prediction for independent and identically distributed samples | Prediction for outcome of intervention | Answering *what-if*, counterfactual questions | Data-driven discovery |
|---|---|---|---|---|---|
| **Mechanistic models** | Pharmacokinetics (PK) and Physically based pharmacokinetic (PBPK) models | yes | yes | yes | maybe |
| **Causal models** | See discussion and examples | yes | yes | yes | maybe |
| **Statistical models** | Cox regression model for survival, statistical tests, (generalised) linear models for *omics* data, support vector machines, random forest, neural networks | yes | no | no | yes |

## Glossary

- **Correlation and causation:** see *Introduction*.
- **Directed acyclic graph (DAG)**: A directed graph without directed cycles. A graph consists of *vertices* and *edges* (also called *arcs*) connecting vertices. If the edges are directed from one vertex to another, the graph is called *directed*. If by following the edge directions, we never form a closed loop (also called a *cycle*), the directed graph is called *acyclic*. See https://en.wikipedia.org/wiki/Directed_acyclic_graph
- **Potential outcomes:** The outcomes of an individual in both the factual and counterfactual worlds. For instance, for an individual participating in the treatment arm of a randomised clinical trials (RCTs), the potential outcomes are the outcomes to both treatment (factual) and control (counterfactual). See https://en.wikipedia.org/wiki/Causal_model#Potential_outcome
- **Counterfactual:** A hypothetical scenario of how the world would have been under circumstances that are counter to the facts. For instance, for an individual participating in the treatment arm of a RCT, the counterfactual is the hypothetical world in which that individual would not have received treatment. See https://en.wikipedia.org/wiki/Causal_model#Counterfactuals
- **Backdoor criterion:** A sufficient condition to find a set of variables Z to deconfound the analysis of the causal effect of a variable X on a variable Y using observational data. A set of confounder variables Z satisfies the *backdoor criterion* if (1) no confounder variable Z is a descendent of X and (2) all *backdoor paths* between X and Y are blocked by the set of confounders. A backdoor path from X to Y is any path from X to Y that starts with an arrow pointing to X. See
https://en.wikipedia.org/wiki/Causal_model#Confounder/deconfounder
- **Frontdoor criterion:** A sufficient condition to find a set of variables Z to measure the causal effect of a variable X on a variable Y using observational data. A set of variables Z satisfies the *frontdoor criterion* if (1) Z intercepts all directed paths from X to Y, (2) there are no unblocked backdoor paths from X to Z, and (3) all backdoor paths from Z to Y are closed by X.
- **Instrument variable:** A variable Z that can be used to determine the causal effect of a variable X on a variable Y using observational data. A variable Z is a valid instrument variable if (1) it induces changes in X, (2) it has no effects on Y other than through the mediation by X, and (3) it is independent of any confounders that affect both X and Y.
- **Estimation trinity:** estimator (the rule), estimate (the numeric value), and the estimand (the quantity to be estimated)
- **Sensitivity analysis (model refutation):** The study of how the uncertainty in the output of a mathematical model can be divided and allocated to different sources of uncertainty in its inputs.
- **Mendelian randomization:** An *instrument variable* method to determine the causal effect of an exposure X on an outcome Y where the instrument is a genetic variant. See *Section 4.*

## Box 1: An anarchy of approaches to causality

Despite a long history of causality research in multiple disciplines, especially statistics, computer science, epidemiology, healthcare, and philosophy of science, there is no consensus way of defining and inferring causality[24]. Impactful approaches for causal inference include the counterfactual outcome framework, the Campbell's framework, Bayesian decision theory, and the Directed Acyclic Graph (DAG) approach[145]. We note that they are not mutually exclusive. In this review we particularly focused on the counterfactual framework and the DAG approach.

A variety of philosophical and scientific approaches have been developed to investigate causality. Below is a list of approaches that mostly impacted the authors:

- The *constant conjunction* by David Hume;
- Path analysis pioneered by Sewall Wright in 1920s[146];
- Potential outcomes proposed by by Jerzy Neyman in 1923, which revealed the possibility of achieving causal inference with randomised trials[147];
- Hill's causality criteria in 1960s[2]
- Ignorability proposed by Donald Rubin in 1970s, known as *unconfoundedness* to epidemiologists and *selection on observables* by economists, which lead to propensity scores[14,148,149];
- Bayesian network, DAGs, and *do* operators proposed between 1980s and 2000s by Judea Pearl[80,150]
- Representation learning for causal inference[135].

- Causal machine learning aiming at individualised causal inference using high-dimensional features to subdivide population, with emerging work in medicine and healthcare[43,151].
- Learning causal relationships represents the next-level challenge that poses a more substantial challenge than causal effect learning[20].
- Challenges of estimation from finite sample, see Shipley (2016)[152]

We close this box with a quote from Richard McElreath: *There is no method for making causal models other than science. There is no method to science other than honest anarchy.*

## Box 2: Emerging software tools for causal inference

- https://github.com/microsoft/dowhy [40]
- Dagitty[153] in R (GNU), http://www.dagitty.net/
- https://github.com/babylonhealth/counterfactual-diagnosis (may be out of scope)
- https://github.com/lingfeiwang/findr
- https://github.com/causal-machine-learning
- https://github.com/uber/causalml
- https://github.com/microsoft/econml
- https://github.com/google/CausalImpact
- https://github.com/IBM/causallib
- https://github.com/uhlerlab/causaldag
- https://github.com/tlverse/tlverse, with tutorial at https://tlverse.org/tlverse-handbook/.